\documentstyle[twoside,rotating,fleqn,psfig,espcrc2]{article}
\title{RHIC data and the multichain Monte Carlo \textsc{DPMJET}-III
 \thanks{Based on a poster submitted to the 17th
 International Conference on Ultra Relativistic Nucleus Nucleus
 Collisions, Jan. 11--17, Oakland, California USA}
 }
\author{
 F.W.Bopp,  J. Ranft\address{Fachbereich Physik, Universit\"at
     Siegen,
 \\
        D--57068 Siegen, Germany},
         R.Engel\address{Forschungszentrum Karlsruhe, Institut f\"ur
		    Kernphysik,\\
        Postfach 3640, D--76021 Karlsruhe, Germany},
        and
					S.Roesler\address{CERN, Geneva,
					Switzerland}
					       }
\begin{document}

%
\newbox\hdbox%
\newcount\hdrows%
\newcount\multispancount%
\newcount\ncase%
\newcount\ncols
\newcount\nrows%
\newcount\nspan%
\newcount\ntemp%
\newdimen\hdsize%
\newdimen\newhdsize%
\newdimen\parasize%
\newdimen\spreadwidth%
\newdimen\thicksize%
\newdimen\thinsize%
\newdimen\tablewidth%
\newif\ifcentertables%
\newif\ifendsize%
\newif\iffirstrow%
\newif\iftableinfo%
\newtoks\dbt%
\newtoks\hdtks%
\newtoks\savetks%
\newtoks\tableLETtokens%
\newtoks\tabletokens%
\newtoks\widthspec%
%
%
\immediate\write15{%
CP SMSG GJMSINK TEXTABLE --> TABLE MACROS V. 851121 JOB = \jobname%
}%
%
%
\tableinfotrue%
\catcode`\@=11
\def\out#1{\immediate\write16{#1}}
%
%
\def\tstrut{\vrule height3.1ex depth1.2ex width0pt}%
\def\and{\char`\&}
\def\tablerule{\noalign{\hrule height\thinsize depth0pt}}%
\thicksize=1.5pt
\thinsize=0.6pt
\def\thickrule{\noalign{\hrule height\thicksize depth0pt}}%
\def\hrulefill{\leaders\hrule\hfill}%
\def\bigrulefill{\leaders\hrule height\thicksize depth0pt \hfill}%
\def\ctr#1{\hfil\ #1\hfil}%
\def\altctr#1{\hfil #1\hfil}%
\def\vctr#1{\hfil\vbox to0pt{\vss\hbox{#1}\vss}\hfil}%
%
%
\tablewidth=-\maxdimen%
\spreadwidth=-\maxdimen%
\def\tabskipglue{0pt plus 1fil minus 1fil}%
%
%
\centertablestrue%
\def\centeredtables{%
   \centertablestrue%
}%
\def\noncenteredtables{%
   \centertablesfalse%
}%
%
%
\parasize=4in%
\long\def\para#1{
   {%
      \vtop{%
         \hsize=\parasize%
         \baselineskip14pt%
         \lineskip1pt%
         \lineskiplimit1pt%
         \noindent #1%
         \vrule width0pt depth6pt%
      }%
   }%
}%
\gdef\ARGS{########}
\gdef\headerARGS{####}
\def\@mpersand{&}
{\catcode`\|=13
\gdef\letbarzero{\let|0}
\gdef\letbartab{\def|{&&}}%
\gdef\letvbbar{\let\vb|}%
}
{\catcode`\&=4
\def\ampskip{&\omit\hfil&}
\catcode`\&=13
\let&0
\xdef\letampskip{\def&{\ampskip}}%
\gdef\letnovbamp{\let\novb&\let\tab&}
}
\def\begintable{
   \begingroup%
   \catcode`\|=13\letbartab\letvbbar%
   \catcode`\&=13\letampskip\letnovbamp%
   \def\multispan##1{
      \omit \mscount##1%
      \multiply\mscount\tw@\advance\mscount\m@ne%
      \loop\ifnum\mscount>\@ne \sp@n\repeat%
   }
   \def\|{%
      &\omit\widevline&%
   }%
   \ruledtable
}
\long\def\ruledtable#1\endtable{%
%
%
%
   \offinterlineskip
   \tabskip 0pt
   \def\widevline{\vrule width\thicksize}
   \def\endrow{\@mpersand\omit\hfil\crnorm\@mpersand}%
   \def\crthick{\@mpersand\crnorm\thickrule\@mpersand}%
   \def\crthickneg##1{\@mpersand\crnorm\thickrule
          \noalign{{\skip0=##1\vskip-\skip0}}\@mpersand}%
   \def\crnorule{\@mpersand\crnorm\@mpersand}%
   \def\crnoruleneg##1{\@mpersand\crnorm
          \noalign{{\skip0=##1\vskip-\skip0}}\@mpersand}%
   \let\nr=\crnorule
   \def\endtable{\@mpersand\crnorm\thickrule}%
   \let\crnorm=\cr
%
%
   \edef\cr{\@mpersand\crnorm\tablerule\@mpersand}%
   \def\crneg##1{\@mpersand\crnorm\tablerule
          \noalign{{\skip0=##1\vskip-\skip0}}\@mpersand}%
   \let\ctneg=\crthickneg
   \let\nrneg=\crnoruleneg
   \the\tableLETtokens
%
%
   \tabletokens={&#1}
%
%
   \countROWS\tabletokens\into\nrows%
   \countCOLS\tabletokens\into\ncols%
%
%
   \advance\ncols by -1%
   \divide\ncols by 2%
   \advance\nrows by 1%
%
%
   \iftableinfo %
      \immediate\write16{[Nrows=\the\nrows, Ncols=\the\ncols]}%
   \fi%
%
%
   \ifcentertables
      \ifhmode \par\fi
      \hbox to \hsize{
      \hss
   \else %
      \hbox{%
   \fi
      \vbox{%
         \makePREAMBLE{\the\ncols}
         \edef\next{\preamble}
         \let\preamble=\next
         \makeTABLE{\preamble}{\tabletokens}
      }
      \ifcentertables \hss}\else }\fi
   \endgroup
   \tablewidth=-\maxdimen
   \spreadwidth=-\maxdimen
}
\def\makeTABLE#1#2{
   {
   \let\ifmath0
   \let\header0
   \let\multispan0
%
%
   \ncase=0%
   \ifdim\tablewidth>-\maxdimen \ncase=1\fi%
   \ifdim\spreadwidth>-\maxdimen \ncase=2\fi%
   \relax
%
   \ifcase\ncase %
      \widthspec={}%
   \or %
      \widthspec=\expandafter{\expandafter t\expandafter o%
                 \the\tablewidth}%
   \else %
      \widthspec=\expandafter{\expandafter s\expandafter p\expandafter r%
                 \expandafter e\expandafter a\expandafter d%
                 \the\spreadwidth}%
   \fi %
   \xdef\next{
      \halign\the\widthspec{%
      #1
      \noalign{\hrule height\thicksize depth0pt}
      \the#2\endtable
%
      }
   }
   }
   \next
}
\def\makePREAMBLE#1{
   \ncols=#1
   \begingroup
   \let\ARGS=0
   \edef\xtp{\widevline\ARGS\tabskip\tabskipglue%
   &\ctr{\ARGS}\tstrut}
   \advance\ncols by -1
   \loop
      \ifnum\ncols>0 %
      \advance\ncols by -1%
      \edef\xtp{\xtp&\vrule width\thinsize\ARGS&\ctr{\ARGS}}%
   \repeat
   \xdef\preamble{\xtp&\widevline\ARGS\tabskip0pt%
   \crnorm}
   \endgroup
}
\def\countROWS#1\into#2{
   \let\countREGISTER=#2%
   \countREGISTER=0%
   \expandafter\ROWcount\the#1\endcount%
}%
\def\ROWcount{%
   \afterassignment\subROWcount\let\next= %
}%
\def\subROWcount{%
   \ifx\next\endcount %
      \let\next=\relax%
   \else%
      \ncase=0%
      \ifx\next\cr %
         \global\advance\countREGISTER by 1%
         \ncase=0%
      \fi%
      \ifx\next\endrow %
         \global\advance\countREGISTER by 1%
         \ncase=0%
      \fi%
      \ifx\next\crthick %
         \global\advance\countREGISTER by 1%
         \ncase=0%
      \fi%
      \ifx\next\crnorule %
         \global\advance\countREGISTER by 1%
         \ncase=0%
      \fi%
      \ifx\next\crthickneg %
         \global\advance\countREGISTER by 1%
         \ncase=0%
      \fi%
      \ifx\next\crnoruleneg %
         \global\advance\countREGISTER by 1%
         \ncase=0%
      \fi%
      \ifx\next\crneg %
         \global\advance\countREGISTER by 1%
         \ncase=0%
      \fi%
      \ifx\next\header %
         \ncase=1%
      \fi%
      \relax%
      \ifcase\ncase %
         \let\next\ROWcount%
      \or %
         \let\next\argROWskip%
      \else %
      \fi%
   \fi%
   \next%
}
\def\counthdROWS#1\into#2{%
\dvr{10}%
   \let\countREGISTER=#2%
   \countREGISTER=0%
\dvr{11}%
\dvr{13}%
   \expandafter\hdROWcount\the#1\endcount%
\dvr{12}%
}%
\def\hdROWcount{%
   \afterassignment\subhdROWcount\let\next= %
}%
\def\subhdROWcount{%
   \ifx\next\endcount %
      \let\next=\relax%
   \else%
      \ncase=0%
      \ifx\next\cr %
         \global\advance\countREGISTER by 1%
         \ncase=0%
      \fi%
      \ifx\next\endrow %
         \global\advance\countREGISTER by 1%
         \ncase=0%
      \fi%
      \ifx\next\crthick %
         \global\advance\countREGISTER by 1%
         \ncase=0%
      \fi%
      \ifx\next\crnorule %
         \global\advance\countREGISTER by 1%
         \ncase=0%
      \fi%
      \ifx\next\header %
         \ncase=1%
      \fi%
\relax%
      \ifcase\ncase %
         \let\next\hdROWcount%
      \or%
         \let\next\arghdROWskip%
      \else %
      \fi%
   \fi%
   \next%
}%
{\catcode`\|=13\letbartab
\gdef\countCOLS#1\into#2{%
   \let\countREGISTER=#2%
   \global\countREGISTER=0%
   \global\multispancount=0%
   \global\firstrowtrue
   \expandafter\COLcount\the#1\endcount%
   \global\advance\countREGISTER by 3%
   \global\advance\countREGISTER by -\multispancount
}%
\gdef\COLcount{%
   \afterassignment\subCOLcount\let\next= %
}%
{\catcode`\&=13%
\gdef\subCOLcount{%
   \ifx\next\endcount %
      \let\next=\relax%
   \else%
      \ncase=0%
      \iffirstrow
         \ifx\next& %
            \global\advance\countREGISTER by 2%
            \ncase=0%
         \fi%
         \ifx\next\span %
            \global\advance\countREGISTER by 1%
            \ncase=0%
         \fi%
         \ifx\next| %
            \global\advance\countREGISTER by 2%
            \ncase=0%
         \fi
         \ifx\next\|
            \global\advance\countREGISTER by 2%
            \ncase=0%
         \fi
         \ifx\next\multispan
            \ncase=1%
            \global\advance\multispancount by 1%
         \fi
         \ifx\next\header
            \ncase=2%
         \fi
         \ifx\next\cr       \global\firstrowfalse \fi
         \ifx\next\endrow   \global\firstrowfalse \fi
         \ifx\next\crthick  \global\firstrowfalse \fi
         \ifx\next\crnorule \global\firstrowfalse \fi
         \ifx\next\crnoruleneg \global\firstrowfalse \fi
         \ifx\next\crthickneg  \global\firstrowfalse \fi
         \ifx\next\crneg       \global\firstrowfalse \fi
      \fi
\relax
      \ifcase\ncase %
         \let\next\COLcount%
      \or %
         \let\next\spancount%
      \or %
         \let\next\argCOLskip%
      \else %
      \fi %
   \fi%
   \next%
}%
\gdef\argROWskip#1{%
   \let\next\ROWcount \next%
}
\gdef\arghdROWskip#1{%
   \let\next\ROWcount \next%
}
\gdef\argCOLskip#1{%
   \let\next\COLcount \next%
}
}
}
\def\spancount#1{
   \nspan=#1\multiply\nspan by 2\advance\nspan by -1%
   \global\advance \countREGISTER by \nspan
   \let\next\COLcount \next}%
\def\dvr#1{\relax}%
\def\header#1{%
\dvr{1}{\let\cr=\@mpersand%
\hdtks={#1}%
\counthdROWS\hdtks\into\hdrows%
\advance\hdrows by 1%
\ifnum\hdrows=0 \hdrows=1 \fi%
\dvr{5}\makehdPREAMBLE{\the\hdrows}%
\dvr{6}\getHDdimen{#1}%
{\parindent=0pt\hsize=\hdsize{\let\ifmath0%
\xdef\next{\valign{\headerpreamble #1\crnorm}}}\dvr{7}\next\dvr{8}%
}%
}\dvr{2}}
\def\makehdPREAMBLE#1{
\dvr{3}%
\hdrows=#1
{
\let\headerARGS=0%
\let\cr=\crnorm%
\edef\xtp{\vfil\hfil\hbox{\headerARGS}\hfil\vfil}%
\advance\hdrows by -1
\loop
\ifnum\hdrows>0%
\advance\hdrows by -1%
\edef\xtp{\xtp&\vfil\hfil\hbox{\headerARGS}\hfil\vfil}%
\repeat%
\xdef\headerpreamble{\xtp\crcr}%
}
\dvr{4}}
\def\getHDdimen#1{%
\hdsize=0pt%
\getsize#1\cr\end\cr%
}
\def\getsize#1\cr{%
\endsizefalse\savetks={#1}%
\expandafter\lookend\the\savetks\cr%
\relax \ifendsize \let\next\relax \else%
\setbox\hdbox=\hbox{#1}\newhdsize=1.0\wd\hdbox%
\ifdim\newhdsize>\hdsize \hdsize=\newhdsize \fi%
\let\next\getsize \fi%
\next%
}%
\def\lookend{\afterassignment\sublookend\let\looknext= }%
\def\sublookend{\relax%
\ifx\looknext\cr %
\let\looknext\relax \else %
   \relax
   \ifx\looknext\end \global\endsizetrue \fi%
   \let\looknext=\lookend%
    \fi \looknext%
}%
%
%
\def\tablelet#1{%
   \tableLETtokens=\expandafter{\the\tableLETtokens #1}%
}%
\catcode`\@=12

\begin{abstract}

Using data from RHIC we are able to systematically improve the two-component
Dual Parton Model (DPM) event generator DPMJET-III. Introducing
percolation parametrized as fusion of chains the model describes
multiplicities and pseudorapidity distributions in nucleus-nucleus
collisions at all centralities. Guided by the d-Au data from RHIC
we {recalibrate} the model to obtain collision scaling
in h-A and d-A collisions.

\end{abstract}
\maketitle

\vspace{-10mm}
\section{Introduction}

Hadronic collisions at high energies involve the production of particles
with low transverse momenta, the so-called \textit{soft} multiparticle
production. The theoretical tools available at present are not sufficient
to understand this feature from QCD alone and phenomenological models
are typically applied in addition to perturbative QCD. The Dual Parton
Model (DPM) \cite{Capella94a} is such a {phenomenological}
model and its fundamental ideas are presently the basis of many of
the Monte Carlo implementations of soft interactions.

One of the most prominent features of the Dual Parton Model has been so
far the independent production and decay of hadronic strings. It seems,
it is just this feature, which has to be modified in the extremely dense
 hadronic systems produced in central heavy ion collisions.

\vspace* {-3mm}
\section{Two--component Dual Parton Model}

\vspace* {-1mm}

\subsection{Hadron--hadron collisions, the Monte Carlo Event Generator
\textsc{PHOJET}} 
\vspace* {1mm}

 \textsc{Phojet}1.12 \cite{Engel95a,Engel95d} is a modern, DPM and
 perturbative QCD based event generator
 describing   hadron-hadron interactions and also
hadronic interactions involving photons.
  \textsc{Phojet} replaces the original \textsc{Dtujet} 
  model \cite{Aurenche92a}, which was the first implementation of this
  combination of  perturbative QCD and the DPM.

%
The DPM combines predictions 
of the large $N_c,N_f$ expansion of QCD \cite{Veneziano74}
and assumptions of duality \cite{Chew78} with
Gribov's reggeon field theory \cite{Gribov67a-e}.
\textsc{Phojet}, being used for the simulation of 
elementary hadron-hadron, photon-hadron and 
photon-photon interactions
with energies greater than 5 GeV, implements the DPM as a two-component model
using Reggeon theory for soft and leading order
perturbative QCD for hard interactions.
Each \textsc{Phojet} collision includes multiple hard and multiple 
soft pomeron 
exchanges, as well as initial
and final state radiation. In \textsc{Phojet} perturbative QCD 
interactions are referred to as hard Pomeron exchange. 
In addition to the model features as described in detail in \cite{PhD-RE},
the version 1.12 incorporates a model for high-mass diffraction
dissociation including multiple jet production and recursive
insertions of enhanced pomeron graphs (triple-, loop- and double-pomeron
graphs). 
\vspace*{1mm}

High-mass diffraction dissociation is simulated as pomeron-hadron or
pomeron-pomeron scattering, including multiple soft and hard
interactions \cite{Bopp98a}. To account for the nature of the pomeron being a
quasi-particle, the CKMT pomeron struc\-tu\-re function \cite{Capella96a}
with a hard gluonic
com\-po\-nent is used. These considerations refer to pomeron exchange
reactions with small po\-me\-ron-momentum transfer, $|t|$. For large
$|t|$ the rapidity gap production (e.g. jet-gap-jet events) is 
implemented on the basis of the color evaporation model \cite{Eboli98a}.

 \begin{figure}[thb]
\begin{center}
 \vspace*{-3cm}\hspace*{-2cm}
\includegraphics[height=10.5cm,width=11cm]{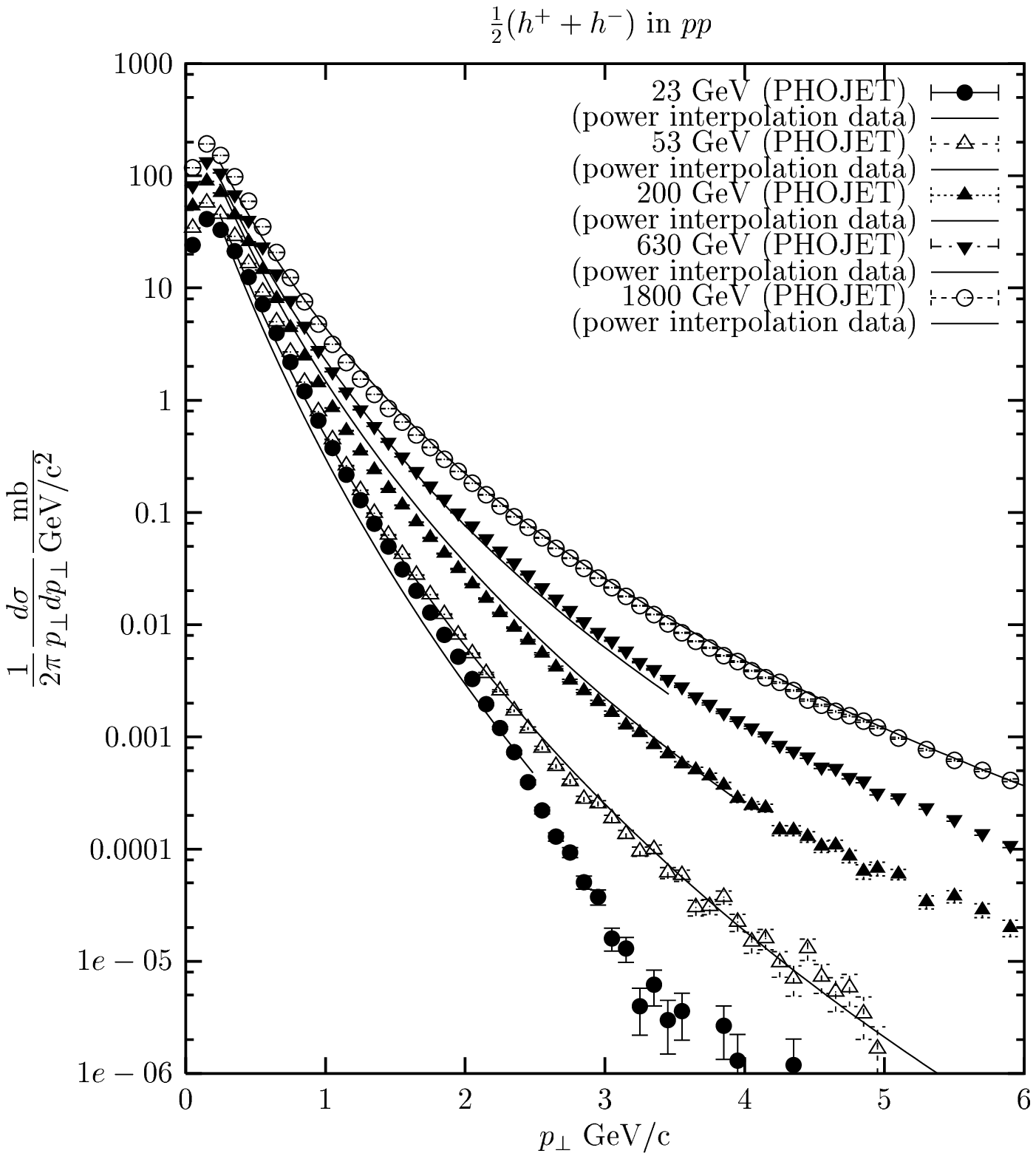}
\end{center}
 \vspace{-3cm}
\caption{\em  Transverse momentum distributions of charged hadrons. 
The results of \textsc{Phojet} (points) are compared to experimental data 
represented by lines, fitted to the data points. The data at $\sqrt{s}$ = 23 
and 53 GeV are from the CERN--ISR\cite{Alper75a}, the data at 200 GeV are from the 
UA1 Collaboration\cite{Arnison82} and the data at 630 and 1800 GeV are from the CDF 
Collaboration\cite{Abe88}.
}
\vspace{-3mm}
\label{fig:pp-pt}
\end{figure} 

\vspace{-1mm}

{The basic building blocks formed in these processes
are color neutral strings. These strings are hadronized in} \textsc{Phojet}
using the Lund model as implemented in \textsc{Pythia}
\cite{Sjostrand01a}. 

\vspace{1mm}
\textsc{Phojet} has been extensively tested against data in hadron--hadron 
collisions \cite{PhD-RE}. Furthermore, in a  number of papers
 LEP Collaborations
compare many features of hadron production in $\gamma$--$\gamma$ collisions
to \textsc{Phojet}, a rather good agreement is usually found.

\vspace{1mm}
In  Fig.\ref{fig:pp-pt} we compare charged hadrons according to 
\textsc{Phojet} against 
practically all data on transverse momentum distributions in p--p and 
$\bar p$--p collisions  from colliders. 
 Please note, the points in this Figure are from the
\textsc{Phojet} Monte Carlo, the data are represented by lines, fits to the
 data points.

\subsection{{Collisions involving nuclei, the Monte Carlo Event
Generator DPMJET-III}}

\vspace{3mm}

 The \textsc{Dpmjet}-III code system \cite{Roesler20001,Roesler20002}
 is a Monte Carlo event 
generator implementing Gribov--Glauber theory for collisions involving
nuclei, for all elementary collisions it uses the  DPM 
as implemented in  \textsc{Phojet}. 
 \textsc{Dpmjet}-III is unique in its wide range of application
 simulating hadron-hadron,
hadron-nucleus,
nucleus-nucleus, photon-hadron, photon-photon and 
photon-nucleus interactions
from a few GeV up to  cosmic ray energies. 

\vspace*{3mm}
 The Gribov-Glauber Multiple Scattering Formalism:
Since its first implementations \cite{Ranft95a,Ranft99b}
 \textsc{Dpmjet} uses
the Monte Carlo realization of the Gribov-Glauber multiple scattering
formalism according to  the algorithms of \cite{Shmakov89} and allows the 
calculation of total, elastic, quasi-elastic and production cross
sections for any high-energy nuclear collision. 
Parameters entering the hadron-nucleon scattering amplitude (total cross
section and slope) are calculated within \textsc{Phojet}.

\vspace*{3mm}
Realistic nuclear densities and radii are used for light nuclei
and Woods-Saxon densities otherwise.

\vspace*{3mm}
During the simulation of an inelastic collision the above formalism samples
the number of ``wounded'' nucleons, the impact parameter of the collision and
the interaction configurations of the wounded nucleons. Individual 
hadron--nucleon interactions are then described by 
\textsc{Phojet} including multiple hard and soft pomeron exchanges, initial
and final state radiation as well as diffraction. 

As a new feature, \textsc{Dpmjet}-III allows the si\-mu\-la\-tion of
enhanced graph cuts in non-diffractive inelastic hadron-nucleus and
nucleus-nucleus interactions. For example, in an event with two wounded
nucleons, the first nucleon might take part in a non-diffractive
interaction whereas the second one scatters diffractively producing only
very few secondaries. Such graphs are predicted by the Gribov-Glauber
theory of nuclear scattering but are usually neglected.
Further features of  \textsc{Dpmjet}-III are a formation zone
intranuclear cascade \cite{Ranft88a} and the implementation of certain
baryon stopping diagrams \cite{Ranft20003}.

\textsc{Dpmjet}-III and earlier versions like \textsc{Dpmjet}-II have been 
extensively tested against data in hadron--nucleus and nucleus--nucleus
collisions\cite{Roesler20001,Roesler20002,Ranft95a,Ranft99a}. The code is used for
 the simulation of cosmic ray showers \cite{Corsika03}.

%
%

\section{Comparing the original \textsc{DPMJET}--III with RHIC data}

\subsection{p--p collisions at 200 GeV 
 }

\begin{figure}[thb]
\begin{center}
\includegraphics[height=7.5cm,width=8cm]{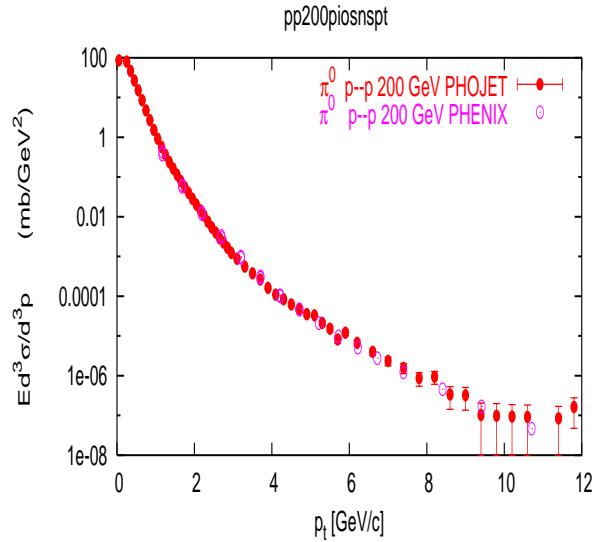}
\end{center}
\vspace{-1cm}
\caption{\em  Transverse momentum distributions of $\pi^0$--mesons
produced in $\sqrt(s)$ = 200 GeV p--p collisions. 
The results of \textsc{Phojet}  are compared to preliminary 
experimental data from the PHENIX--Collaboration \cite{PHENIX02a}.
}
\vspace{-7mm}
\label{fig:pppio-pt}
\end{figure}

In Fig.\ref{fig:pppio-pt} we compare the 
preliminary $\pi^0$ transverse momentum distribution
in p--p collisions at $\sqrt s$ = 200 GeV from PHENIX \cite{PHENIX02a}
to \textsc{Phojet} 
and find excellent agreement up to transverse momenta of about
$p_{\perp}$ = 10 GeV/c. 
 
\vspace{-15mm}

\subsection{Au--Au collisions at 130 GeV in 
requiring Percolation }

The groups at Lisboa \cite{Dias2000a,Dias2000b,Dias2001a} 
and Santiago de Compostela \cite{Braun2002a} were the
first to point out, that the multiplicities measured at RHIC are
significantly lower than predicted by conventional multi--string
models. A new process is needed to lower the multiplicity in
situations with a very high density of produced hadrons like in central
nucleus--nucleus collisions. 
{The
percolation process, which leads with increasing density to more and
more fusion of strings, is one such mechanism} \cite{Braun99,Braun2000a}.

Using the original \textsc{Dpmjet}--III with enhanced baryon stopping
and a centrality of 0 to 5 \% we compare to some multiplicities measured
in Au--Au collisions at RHIC for $\sqrt{s}_{NN}= 130$ GeV. 

{\begin{center} \small
{}\begin{tabular}{|c|c|c|}
\hline 
&

{$N_{ch}$}
&

{$dN_{ch}/d\eta|_{\eta=0}$}
\tabularnewline
\hline
\hline 
\textsc{{Dpmjet}}{--III}&

{6031}&
{968}\tabularnewline
\hline 
{BRAHMS\cite{BRAHMS01a}}&

{3860 $\pm$ 300}&
{553 $\pm$ 36}\tabularnewline
\hline
{PHOBOS \cite{PHOBOS02a}}&

&
{613 $\pm$ 24 }\tabularnewline
\hline
\hline 
{PHENIX \cite{PHENIX00a}}&

&
{622 $\pm$ 41}\tabularnewline
\hline
\end{tabular}  \end{center}
}

\vspace*{2mm}
{This} comparison of \textsc{Dpmjet}--III with multiplicity
data from RHIC {confirms: there is} a new mechanism
needed to reduce $N_{ch}$ and $dN_{ch}/d\eta|_{\eta=0}$ in situations
with a produced very dense hadronic system. Therefore, in the next
Section we will introduce percolation and chain fusion into \textsc{Dpmjet}--III.
     
\vspace*{2mm}

\section{{Percolation of hadronic strings in the modified} \textsc{DPMJET}{--III}}

We consider only the percolation  {with} fusion of {complete}
soft chains, these are chains where the
transverse momenta of both chain ends is below a certain cut--off
$p_{\perp}^{fusion}$. At the moment we use $p_{\perp}^{fusion}$ = 2
GeV/c. The condition of percolation is, that the chains overlap in
transverse space. We calculate the transverse distance of the chains $L$
and $K$ $R_{L-K}$ and allow fusion of the chains for  $R_{L-K} \leq
R^{fusion}$. At the moment we use $ R^{fusion}$ = 0.75 fm. This
procedure is certainly the simplest approximation. A better procedure
could be to introduce $p_{\perp}$ dependent transverse dimensions of the
chains and then check for percolation.

The chains in \textsc{Dpmjet} are fragmented using the Lund code \textsc{Jetset}
as available inside the \textsc{Pythia} code \cite{Sjostrand01a}.
Only the fragmentation of color triplet--antitriplet chains is available
in \textsc{Jetset}, however fusing two arbitrary chains could result
in chains with other colors. Therefore at the moment we select only
chains for fusion, which again result in triplet--antitriplet chains.
Examples for the fusion of two chains are:

\noindent
(i)A $q_{1}-\bar{q}_{2}$ plus a $q_{3}-\bar{q}_{4}$ chain 
\vspace*{-3mm}
\begin{flushright}become a $q_{1}q_{3}-\bar{q}_{2}\bar{q}_{4}$
chain.\end{flushright}
\vspace*{-3mm}
(ii)A $q_{1}-q_{2}q_{3}$ plus a $q_{4}-\bar{q}_{2}$ chain 
\vspace*{-3mm}
\begin{flushright}become a $q_{1}q_{4}-q_{3}$ chain.\end{flushright}
\vspace*{-3mm}
Examples for the fusion of three chain are:

\noindent
(iii)A $q_{3}-q_{1}q_{2}$, a $q_{4}-\bar{q}_{1}$ plus a $\bar{q}_{3}-q_{5}$
chain \vspace*{-3mm}
\begin{flushright}become a $q_{4}-q_{2}q_{5}$ chain.\end{flushright}
\vspace*{-3mm}
(iv)A $q_{4}-\bar{q}_{1}$, a $q_{5}-\bar{q}_{3}$ plus a $\bar{q}_{5}-q_{1}$
chain 
\vspace*{-3mm}
\begin{flushright}become a $q_{4}-\bar{q}_{3}$ chain.\end{flushright}
\vspace*{-3mm}

The expected results of these transformations are a decrease of the
number of chains. Even if the fused chains have a higher energy than
the original chains, the result will be a decrease of the hadron 
multiplicity $N_{hadrons}$. In reaction (i) we observe new diquark and
anti--diquark chain ends. In the fragmentation of these chains we expect
baryon--antibaryon production anywhere in the rapidity region of the
collision. Therefore, (i) helps to shift the antibaryon to baryon ratio
of the model into the direction as observed in the RHIC experiments.

 \begin{figure}[thb]
\vspace{-8mm}
 \centerline{\psfig{file=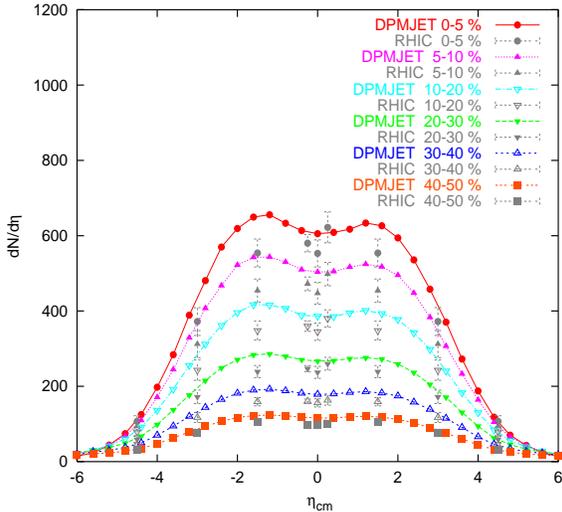,width=7.8cm,height=69mm}}
 \vspace*{-11mm}
 \caption{Pseudorapidity distributions of charged hadrons in Au--Au
 collisions at $\sqrt{s}_{NN}$= 130 GeV for centralities 0--5 \% up to
 40--50 \%. 
 The points {(}with rather small error bars 
{)} are from the \protect{}\textsc{Dpmjet}--III Monte Carlo with
chain fusion as described in the text. {With two exceptions}
data points are from the BRAHMS Collaboration \cite{BRAHMS01a}. {The
$\eta$=0.0 points displayed at $\eta$=0.25 are from PHENIX \cite{PHENIX00a}
and the $\eta$=0.0 points displayed at $\eta$=-0.25 are from 
PHOBOS} \cite{PHOBOS02a}. \protect\label{dpmfus130etach}
 }
\vspace*{-9mm}
\end{figure}

\subsection{ d-Au collisions in 
{\sc Dpmjet}-III
 }
       
\begin{figure}[thb]
\vspace{-11mm}
\begin{center}
\includegraphics[height=8cm,width=8cm]{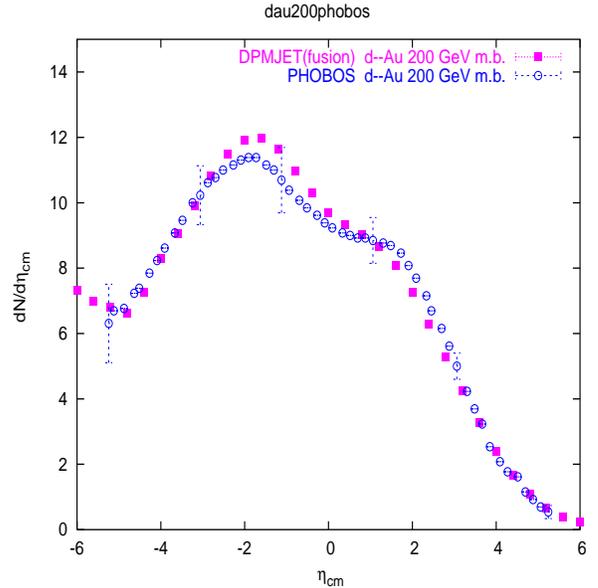}
\end{center}
\vspace{-13mm}

\caption{Pseudorapidity distribution of charged hadrons produced in minimum
bias $\sqrt{s}$ = 200 GeV d--Au collisions. The results of \textsc{Dpmjet}
with percolation {with} fusion of chains are compared
to experimental data from the PHOBOS--Collaboration \cite{PHOBOSdau}.
We give at some pseudorapidity values the systematic errors as estimated
by the experimental collaboration. }

\vspace{-7mm}
\label{fig:etacmdau200}
\end{figure}

{d-Au collisions are the first example where percolation with fusion of
chains is found to be important.} Pseudorapidity
distribution of charged hadrons produced in minimum bias $\sqrt{s}$
= 200 GeV d--Au collisions were measured at RHIC by the PHOBOS--Collaboration
\cite{PHOBOSdau}. 
In Fig.\ref{fig:etacmdau200} we compare the PHOBOS
data to \textsc{Dpmjet}-III calculations. Using \textsc{Dpmjet}-III
without percolation {with} fusion of chains we find
the \textsc{Dpmjet} distribution above the experimental data outside
the systematic errors. Using \textsc{Dpmjet}-III {with
fusion} of chains we find the \textsc{Dpmjet} distribution within
the systematic errors.

Several RHIC experiments (see for instance 
 \cite{PHENIX02dau}) find in d-Au collisions at large $p_{\perp}$ a 
 nearly perfect collision scaling for $\pi^0$ production. 
(Collision scaling means $R_{AA} \approx 1.0$. )
The $R_{AA}$ ratios are defined as follows:
\begin{equation}
R_{AA} =
\frac{\frac{d^2}{dp_{\perp}d\eta}N^{A-A}}{N^{A-A}_{binary} \cdot
\frac{d^2}{dp_{\perp}d\eta}N^{N-N}}
\end{equation}
Here $N^{A-A}_{binary}$ is the number of binary Glauber collisions in
the nucleus--nucleus collision A--A. 

\begin{figure}[thb]
\vspace*{-11mm}
\begin{center}
\includegraphics[height=6.0cm,width=8cm]{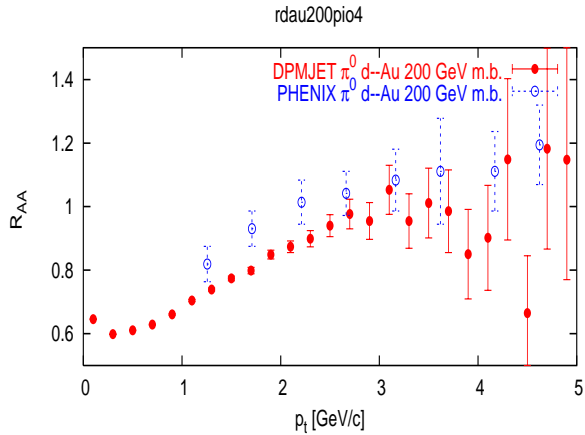}
\end{center} \vspace{-11mm}
\caption{\em $R_{AA}$ ratio  of $\pi^0$--mesons
produced in $\sqrt{s}$ = 200 GeV d--Au collisions. 
The results of the modified \textsc{Dpmjet}  are compared to  
experimental data from the PHENIX--Collaboration \cite{PHENIX02dau}.
}
\vspace{-7mm}
\label{fig:Raadau200}
\end{figure}

 \textsc{Dpmjet}--III  in its original form gave for $\pi^0$ production in
 d+Au collisions strong deviations from collision scaling ($R_{AA}
 \approx$ 0.5 at large $p_{\perp}$). The reason for this was in the
 iteration procedure to sample the multiple  collisions
 in \textsc{Dpmjet}: some soft and hard collisions were rejected by this
 iteration procedure. 
 Using a {reordered}
iteration procedure it was possible to obtain a nearly perfect collision
scaling, see Fig. \ref{fig:Raadau200}.

\subsection{ Au-Au collisions in 
{\sc Dpmjet}-III                 }

Our \textsc{Dpmjet}--III studies on hadron production in central Au--Au
collisions are still in a rather preliminary status. \textsc{Dpmjet}--III
in its original form gave for $\pi^{0}$ production in central Au+Au
collisions strong deviations from collision scaling. 
{In
contrast to the low $p_{\perp}$ behavior,} the $\pi^{0}$ production
at large $p_{\perp}$ was rather similar to the STAR--data\cite{Adlerauaupio}.
{Central to that behavior} was that in the iteration
procedure to implement the multiple collisions in \textsc{Dpmjet}--III
some soft and hard c{ollisions were rejected. }

{T}he RHIC data on hadron production in d--Au collisions
made it clear, that this iteration procedure had to be changed in
such a way that no essential deviations from collision scaling occur
in d--Au collisions. The same corrections to the iteration procedure
{then also} influence Au--Au collisions i{n
a} way, that the deviations from collision scaling are reduced. Hence
we have to recalculate all hadron production at large $p_{\perp}$
in Au--Au collisions and subsequently we have to introduce nuclear
modifications to parton distributions and interactions of the scattering
partons in the dense medium. These modifications have not yet {been}
implemented so far.

\begin{figure}[thb]
\begin{center}
\includegraphics[height=8.0cm,width=8cm]{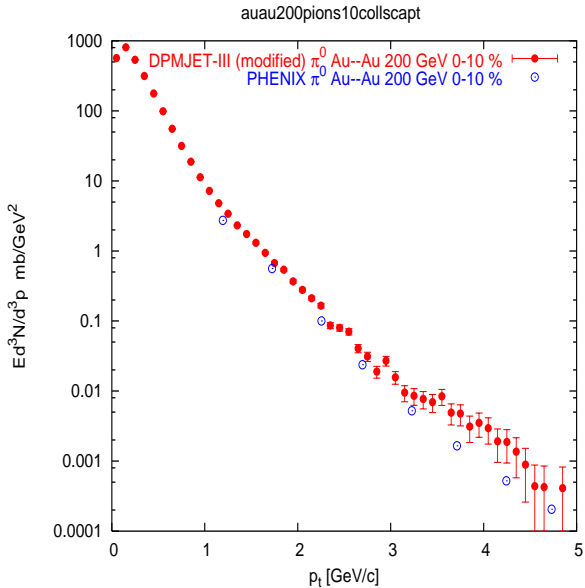}
\end{center}
\vspace{-1cm}
\caption{\em  Transverse momentum distribution  of $\pi^0$--mesons
produced in central $\sqrt{s}$ = 200 GeV Au--Au collisions. 
The results of the modified \textsc{Dpmjet}--III  are compared to  
experimental data from the PHENIX--Collaboration \cite{Adlerauaupio}.
}
\vspace{-7mm}
\label{fig:auau200piocollscapt}
\end{figure}

The only thing we can present at the moment is a recalculation of large
$p_{\perp}$ $\pi^0$ production with  {\sc Dpmjet}--III with full
percolation and chain fusion and with the kinematical corrections to
change the iteration procedure. In Fig.\ref{fig:auau200piocollscapt} we
present such {\sc Dpmjet}--III results and compare them to the RHIC
data\cite{Adlerauaupio}. As expected, at larger $p_{\perp}$ {\sc
Dpmjet}--III produces a  $p_{\perp}$ distribution above the data.

\section{ Elliptic flow in 
{\sc Dpmjet}-III
 }
       
The elliptic flow $v_2$ is defined as a Fourier coefficient of the
single particle distribution \cite{Voloshin96,Ollitrault92}
$$
v_2 = <cos{2(\phi  - \Phi_R)}>.
$$
$\Phi_R$ is the orientation of the reaction plane and the angular
brackets denote an average over many particles belonging to some
phase--space region and over many collisions having approximately the
same impact parameter. The elliptic flow $v_2$ is a sensitive probe of
the dense matter produced in high--energy nucleus--nucleus collisions.
Data from RHIC on $v_2$ were published by most RHIC Collaborations
\cite{Adcox02,Back02,Ackermann01,Adler02}.

Quite a number of different methods have been proposed to extract the
elliptic flow from data or from Monte Carlo models, it is not the
purpose of this contribution to give a  overview over these methods. Let us
only mention two methods, which we use to analyze elliptic flow from
{\sc Dpmjet}-III. One method uses {in
its simplest form two-particle cumulants \cite{Borghini01}. Th}e
second me\-th\-od determines flow using Lee--Yang zeroes \cite{Bhalerao03}.

We concentrate on Au--Au collisions at $\sqrt{s}_{NN}$ = 130 GeV, for
which 
experimental data for the integrated flow and for $p_{\perp}$ dependent
flow were presented in Ref.~\cite{Adler02}. In a multichain model
like \textsc{Dpmjet}-III with non-interacting hadronic strings fragmenting
into hadrons and for large secondary multiplicities, we expect of course
a vanishing elliptic flow. But there are two effects in \textsc{Dpmjet}-III
which are expected to change this picture. The first effect is {connected
to the percolation} of hadronic strings in \textsc{Dpmjet}-III. An 
interaction of hadronic strings {can be expected to
also} introduce  a force between strings yielding some genuine elliptic flow. The second effect are
jets and minijets in \textsc{Dpmjet}-III. The number of large $p_{\perp}$
jets and minijets at RHIC energies is not large, the jets produce
significant azimuthal fluctuations, which leads to non-vanishing $v_{2}$
especially at large $p_{\perp}$.

\begin{figure}[thb]
\begin{center}
\includegraphics[height=8cm,width=8cm]{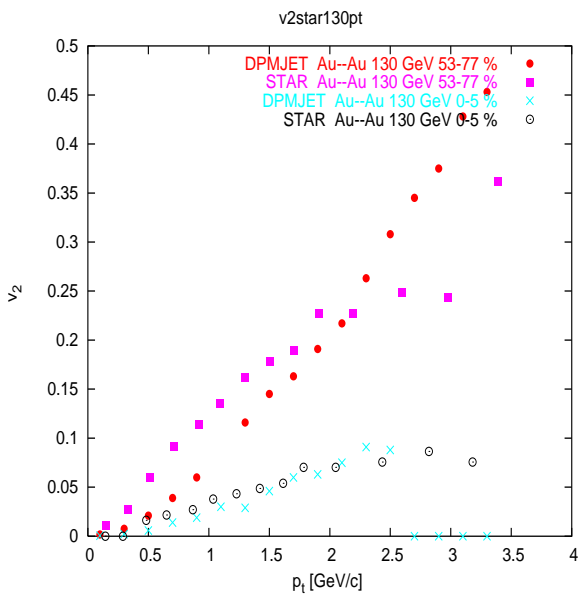}
\end{center}
\vspace{-1cm}
\caption{Elliptic flow \protect$v_2$ as function 
of \protect$p_{\perp}$ in 130 GeV
Au--Au collisions at RHIC. We compare for charged hadrons the  
{\sc Dpmjet}-III results for two centralities to experimental data from
the STAR--Collaboration \protect\cite{Adler02}
}
\vspace{-7mm}
\label{fig:starflow130}
\end{figure}

In Table 1 we collect for two centralities (and the pseudorapidity
acceptance of the experiment) some integrated flow values calculated
with both methods mentioned above. In Fig.\ref{fig:starflow130} we
compare the $p_{\perp}$ dependent elliptic flow obtained from  {\sc
Dpmjet}-III using the two--particle cumulant method\cite{Borghini01} for
two different centralities to STAR data \cite{Adler02} from RHIC.


Let us discuss {the results of the c}omparisons in
Table 1 and Fig.\ref{fig:starflow130}.

 (i)The integrated flow values calculated from {\sc Dpmjet}-III 
 with methods I and II agree
 reasonably well, however the spurious flow (column III of Tab.~1) 
calculated for method II
 indicates, that method II using Lee--Yang zeroes is not able to
 calculate the flow reliably. Also we find, that we are not able to
 calculate a significant $p_{\perp}$ dependent flow using method II.

(ii)The integrated flow values  in Table 1 
calculated from {\sc Dpmjet}-III are significantly smaller than the
experimental data from STAR, also we see in Fig.\ref{fig:starflow130}
the $p_{\perp}$ dependent flow at low $p_{\perp}$ values (which dominate
the particle
production) is significantly smaller than the experimental data from
STAR. We conclude: the flow introduced in the model by the percolation
and fusion of chains is not enough to explain the data.
%

\begin{table}\caption{Integrated flow: $v_{2}$ results obtained from \textsc{Dpmjet}-III
using two different methods (I: using two particle cumulants\cite{Borghini01};
II: using Lee--Yang zeroes \cite{Bhalerao03}; III: gives the spurious
flow according to method II) to STAR data \cite{Adler02} from RHIC.  }
\vspace{10mm}
{\small \begintable
{\tiny Centrality}|{\tiny particles}|~I~|~II~| ~III~|~STAR~\cr
 53--77 \%|charged |0.0107 |0.0140 |0.047 |0.07 \cr
 53--77 \%| $\pi^{\pm}$|0.0182 |0.0135 |0.051 | \cr
 53--77 \%| $K^{\pm}$| 0.060 |0.037 |0.164 | \cr
 0--5 \%|charged |0.00685 |0.00644 |0.0129 |0.02 \cr
 0--5 \%|$\pi^{\pm}$ | 0.00595 |0.00693 |0.0139 | \cr
 0--5 \%|$K^{\pm}$ |0.0188 |0.0215 |0.0419 | \endtable }
\end{table}  

(iii)We find in Fig.\ref{fig:starflow130} that  {\sc Dpmjet}-III at
large $p_{\perp}$ values shows larger flow than the data, 
this flow results from the jets and minijets as discussed above. This
flow component in the experimental data is significantly smaller, we
know that large  $p_{\perp}$ jets in the Au--Au data are 
strongly suppressed, the present version of  {\sc Dpmjet}-III does not
contain any mechanism for jet quenching.

 \vspace{5mm}

\section{ Summary 
 }
 
 The data obtained at RHIC are extremely useful to
 improve hadron production models like {\sc Dpmjet}-III.

 Of particular importance are data on hadron production in p--p
 collisions, d-Au collisions and peripheral Au--Au collisions, in all of
 these collisions (unlike central Au-Au collisions) we do not expect any
 change in the reaction mechanism, which could not be accommodated into
 the mechanisms as implemented in {\sc Dpmjet}--III.

 Indeed, comparing {\sc Dpmjet}--III to RHIC data we find two important
 corrections to be applied to {\sc Dpmjet}--III, which otherwise do not
 completely change the independent chain fragmentation model: 
 (i) Percolation and
 fusion of chains, the data from RHIC allow to determine the amount of
 percolation to be implemented into {\sc Dpmjet}--III. (ii) Collision
 scaling of large $p_{\perp}$ hadron production in d--Au collisions: The
 data indicate that we have 
 to change the iteration procedure (of the selection of all soft and
 hard chains in nuclear collisions) in such a way, 
 that collision scaling
 is obtained.

 
 \bibliographystyle{prsty}

 \bibliography{dpm11}
 

\end{document}